\documentclass{article}
\usepackage{makeidx}  
\usepackage{color}
\usepackage{cite}
\usepackage{graphicx}
\usepackage{subfigure}

\begin{document}
\title{On the usefulness of information hiding techniques for wireless sensor networks security}

\author{Rola Al-Sharif, Christophe Guyeux, Yousra Ahmed Fadil,\\ Abdallah Makhoul, and Ali Jaber}
%
%

\maketitle              

\begin{abstract}        
A wireless sensor network (WSN) typically consists of base stations and a large number of wireless sensors. The sensory data gathered from the whole network at a certain time snapshot can be visualized as an image. As a result, information hiding techniques can be applied to this ``sensory data image''. Steganography refers to the technology of hiding data into digital media without drawing any suspicion, while steganalysis is the art of detecting the presence of steganography. This article provides a brief review of steganography and steganalysis  applications for wireless sensor networks (WSNs). Then we show that the steganographic techniques are both related to sensed data authentication in wireless sensor networks, and when considering the attacker point of view, which has not yet been investigated in the literature. Our simulation results show that the sink level is unable to detect an attack carried out by the nsF5 algorithm on sensed data.
\end{abstract}

\section{Introduction}

Wireless sensor network (WSN) typically consists of base stations and a number of wireless sensors~\cite{bgmp12:ij}. Sensors are usually small in size, have limited computing capabilities, communicate wirelessly and are powered by small batteries. These sensors are often scattered in a sensor field. Data from the sensor field are collected and sent to a base station. The base station then sends the data to the end users for analysis and strategic decisions. Base stations usually have unlimited power, sufficient memory, powerful processors and a high bandwidth link, in comparison to other sensor nodes.

WSNs are used in many fields. For example, like in military applications for monitoring friendly forces, battlefield surveillance, biological attack detection, troop coordination, and battle damage assessments. In environmental applications, sensors can be used to detect and monitor environmental changes like tracking oil pollution.
Data being transmitted are vulnerable to external or internal attacks, such as forgery, tampering, replay and selective forwarding. Data integrity is a core requirement for secure sensor data in WSN. False or malicious data would result in incorrect decisions and potentially financial losses. For instance, an intruder could insert hidden code or a message in the network with intent to do harm. On the battlefield this could result in misinformation to troops that could put them in harm's way. In a health care application, an intruder could insert code that result in the relay of false information about a patient who is being monitored. Such incidents could result in lack of confidence in the security of WSNs.

On the other hand, in an era of rising security concern, steganography (the art of secret communication) and steganalysis (attacks against steganography to discover hidden messages) have taken an increased importance. The information hiding (IH) techniques are reputed to meet both legal and illegal interests. For example, civilians may use it for protecting privacy while terrorists may use it for spreading terroristic information.

In this article, we show that this claim holds too in WSN context, by illustrating the fact that, on the attacker side, steganographiers can be used to manipulate data without being detected, while steganalyzers are useful for the sink to detect any malfunctioning behavior. Our aim is to prove that IH techniques can enrich the collection of tools useful for either guaranteeing or attacking wireless sensor networks. Therefore, we start our work by reviewing existing proposals using IH techniques for WSN security, then we propose a new context of applications with concrete examples and validated via simulation results.

The organization of this paper is as follows. In the next section, we will remind definition of steganography and steganalysis. In Section~\ref{sec:appStegana},  a practical application of steganalysis for detecting an abnormality in a network is presented. Then, in Section~\ref{sec:appStegano}, we present a concrete application of steganography to perform an attack on a network. This article ends by a conclusion section, in which the contribution is summarized and intended future work is detailed.

\section{Steganography and Steganalysis}

\subsection{Steganography}

Since the rise of the Internet one of the most important factors of information technology and communication has been its security. To do so, encryption has been developed as a technique for securing communications. Many different methods have been developed to encrypt and decrypt data in order to keep the message secret. But it may not be enough to keep the contents of a message secret, it may also be necessary to keep the existence of the message secret. The technique used to implement this is called steganography~\cite{Morkel05,bcg11:ij}.

Steganography is different from cryptography. Where cryptography focuses on keeping the contents of a message secret, steganography focuses on hiding the fact that a secret message exists. 
Note that (1) the strength of steganography is amplified by combining it with cryptography: first the secret message is encrypted and then it is embedded into other cover content.
(2) Images are the most popular carrier files for steganography, because of 
the way images are stored creates a great amount of redundant space, which is the ideal place to hide information. Hiding information is done through a variety of algorithms, mainly of them being based on bit-level.

\subsection{Image Steganography}


Researches mainly concentrate on hiding data in gray-scale or color images. Since the luminance component of a color image is equivalent to a gray-scale image, we focus on the steganography for gray-scale images. Besides, it is generally considered that gray-scale images are more suitable than color images for hiding data because the disturbance of correlations between color components may easily reveal the trace of embedding~\cite{Morkel05, Chanu12}.

\begin{description}
\item[Spatial steganography.]
The common ground of spatial steganography is to directly change the image pixel values for hiding data. The embedding rate is often measured in bit per pixel (bpp). One of the most reputed tool of this kind is the so-called HUGO steganographier~\cite{GulK11}.
\item[JPEG steganography.]
JPEG is one common format of the images produced by digital cameras, scanners, and other photographic image capture devices. Therefore, hiding secret information into JPEG images may provide better camouflage. Most of the steganographic schemes embed data into the non-zero alternate current (AC) discrete cosine transform (DCT) coefficients of JPEG images. As a result, the embedding rate of JPEG steganography is often evaluated in bit per non-zero AC DCT coefficient (bpac). Currently, the 
best frequency domain steganographier is the nsF5 algorithm~\cite{nsF5}, described in a next section. 
\end{description}

\subsection{Steganalysis}

Because steganography is a technology that enables users to hide messages from unintended recipients, it can also be used by criminals to hide messages from authorities. None of these have been provenly used, but the fact that these possibilities exist makes it necessary to research methods for detecting steganography. Such methods are called steganalysis~\cite{Singh12}.
Indeed, for
steganographic algorithms sometimes leave a signature in the 
stego-content, which can be detected by \textit{ad hoc} artificial intelligence tools.

More precisely, steganalysis can be regarded as a two-class pattern classification problem that aims to determine whether a testing medium is a cover medium or a stego one. 
Steganalysis can be realized following
either
specific methods or universal ones. A specific steganalytic method fully utilizes the knowledge of a targeted steganographic technique and may only be applicable to such a kind of steganography. A universal steganalytic method can be used to detect several kinds of steganography. Conversely, universal methods do not require the knowledge of the details of the embedding operations. Therefore, it is also called blind method~\cite{Chanu12, Singh12}.




\section{Relations between images and WSN}

\label{sec:appStegana}


The key idea formerly presented in \cite{Zhang:2008} is to visualize the sensory data gathered from the whole network at a certain time snapshot as an image, in which every sensor node is viewed as a pixel with its sensory reading representing the pixel's intensity. As a result, information hiding techniques can be applied to this ``sensory data image''. Specifically, they adopt direct spread spectrum sequence (DSSS) based watermarking to balance energy consumption in the network with asymmetric resources. With a simple mathematical operation (addition), each sensor node can embed part of the whole watermark into its sensory data, while leaving the heavy computation load from watermark detection at the sink. Once the aggregated and watermarked data reach the sink, this latter is able to verify the existence of the watermark and hence the authenticity of the data.

Authors of this aforementioned article adopt existing image compression schemes as aggregation functions, to reduce network load while retaining the most informative part of the data.
Recall that for frequency domain compression like JPEG, an image is first divided into a number of non-overlapping blocks. Then, roughly 
speaking, a linear transform such as (DCT) or (DWT), is applied to each block to transform the data into frequency domain, and smallest 
coefficients of this transform are
set to 0. Therefore it is possible
to slightly alter informative 
frequential coefficients, in order to
embed a secret message in them, which
still remains after compression.

The proposal of~\cite{Zhang:2008} can
thus be summarized as follows.
Based on the block size (system parameter), a cluster of sensors is first divided into blocks, in each of which a DCT is performed by the cluster head.
Once the aggregated and watermarked data reaches the sink, the sink is able to verify the existence of the watermark and hence the authenticity of the data.

\subsection{Steganalyzers as malfunctioning detectors}

Let us firstly recall a few words about steganalyzers, that is
tools designed to detect the presence of hidden information
into a given innocent looking cover image.

The oldest steganalytic technique is visible detection, which 
include human observers detecting minute changes between 
a cover file and a stego one. 
For palette-based images, if the embedded file was 
inserted without first ordering the cover file palette 
according to its colors, then dramatic color shifts can be 
found in the stego file. Additionally, since many 
steganographic tools take advantage of close colors or 
create their own close color groups, many similar 
colors in an image palette may make the image  
suspect~\cite{2001:IHS}. By filtering images as described 
by Westfield and Pfitzmann in~\cite{Westfeld00}, the 
presence of an embedded file can become obvious to the human observer.

Steganalysis can also involve the use of statistical techniques. 
By analyzing changes in an image's close color pairs, the 
steganalyst can determine if LSB substitution was used. 
Close color pairs consist of two colors whose binary 
values differ only in the LSB. The sum of occurrences of 
each color in a close color pair does not change between 
the cover file and the stego file \cite{Westfeld00}. This 
fact, along with the observation that LSB substitution merely 
flips some of the LSBs, causes the number of occurrences 
of each color in a close color pair in a stego file 
to approach the average number of occurrences for that 
pair~\cite{2001:IHS}. Determining that the number of 
occurrences of each color in a suspect image's close 
color pairs are very close to one another gives a 
strong indication that LSB substitution was used 
to create a stego file \cite{Westfeld00}.

Fridrich and others proposed a steganalytic technique 
called the RQP method. It is used on color images with 
24-bit pixel depth where the embedded file is encoded 
in random LSBs. RQP involves inspecting the ratio 
between the number of close color pairs and all pairs of colors. This ratio is calculated on the suspect image, a test message is embedded, and the ratio is calculated again. If the initial and final ratios are vastly different then the suspect image was likely clean. If the ratios are very close then the suspect image most likely had a secret message embedded in it~\cite{fridrich}.

These statistical techniques benefit from the fact that the embedding process alters the original statistics of the cover file and in many cases these first order statistics will show trends that can raise suspicion of steganography \cite{fridrich, Westfeld00}. However, steganographic tools such as OutGuess \cite{Provos:2001} are starting to maintain the first-order statistics during the embedding process. Steganalytic techniques using sensitive higher-order statistics have been developed to counter this covering of tracks \cite{Farid:2001, Fridrich02}.

\section{Using steganographic techniques for wireless sensor networks}
\label{sec:appStegano}

\subsection{Digital Watermarking and WSN}

Current technologies allow validation during data transit through the WSN, but stop after the data reaches its destination (a specific
node or the sink). One of the challenges with these technologies is to 
preserve the source of the data
once they leave the WSN. 
Therefore, it needs to be ensured that the data source is identifiable and the data is valid. Sensors are susceptible to various types of attack, such as data modification, data insertion and deletion, or even physical capture and sensor replacement. Hence, security becomes an important issue with WSNs. Traditional algorithms are used for securing data transmission between sensor nodes. However these algorithms need millions of multiplication instructions to perform operations, and cannot efficiently protect the copyright of the valuable sensor data. Digital watermarking techniques are one of the effective choices to overcome this challenge: a watermark is added as a second line of defense, to ensure that the data is valid.

Two manners to embed a piece of information (the watermark) into the data stream are possible.
The embedding can be achieved in such a way that any change or tampering with the original data would corrupt the watermark: this type of digital watermarking is called fragile watermarking. 
Conversely, in robust watermarking, the added information cannot be removed without destroying the
cover information. Fragile watermarking is useful to detect any attempt to tamper with sensed data, while 
robust watermarking serves when data authentication is required (note that cryptography provides no protection after the content is decrypted).
Watermarking algorithms are much lighter and thus require less battery power and processing capabilities than cryptographic-based algorithms. Another advantage for watermarking-based algorithms is that the watermark is embedded directly into the sensor data: the payload does not increases. 

\subsection{Another application of steganographic tools for WSNs}

\subsubsection{Using steganographiers to achieve WSN attacks}

Steganographic techniques are not only related to sensed data authentication
in wireless sensor networks. Another application, which has not yet been investigated
in the literature, is to consider the attacker point of view. 

Let us suppose that this
latter desires to manipulate the data in such a way that this manipulation cannot
be detected in sink side. For large scale networks, such a detection and the surveillance
of data provided by the wireless sensor network cannot be
achieved manually. It necessitates ad hoc techniques for manipulating big data, such
as data mining, information theory, or artificial intelligence. Similar techniques have
been deployed for detecting artificial manipulations of images or videos, 
steganalyzers being among the best tools currently available. Thus it is reasonable
to consider, as a first approximation, that the sink either embeds such a tool, or at
least uses another device having a similar behavior.

A way to achieve such an attack is thus to consider the parallel presented in a previous 
section, between images and sensory data gathered from the whole network at a certain time snapshot,
and to map the modifications of the area following the
locations designed by the steganographic tools under consideration.
Doing so using an up-to-date steganographier like nsF5 (see below) for achieving, for instance,
an intrusion on an area under video-surveillance, leads to slight 
modifications of the sensed data very difficult to detect, and the adversary
can consequently hope to achieve his attack without being detected.

In the following, we will simulate an attack on sensed data using nsF5 algorithm, 
and we will prove that even up-to-date steganalyzers are not able to detect 
any abnormal behavior in the data provided by the WSN. Let us first recall how 
the nsF5 works.

\subsubsection{Presentation of nsF5}
The nsF5 algorithm \cite{nsF5} extends the F5 algorithm \cite{F5}. Let us first have a closer look on this latter.
First of all, as far as we know, F5 is the first steganographic approach that solves the problem of remaining unchanged a part (often the end) of the file. To achieve this, a subset of all the least significant bits LSB is computed thanks to a pseudorandom number generator seeded with a user defined key. Next, this subset is split into blocks of x bits. The algorithm takes benefit of binary matrix embedding to increase it efficiency. Let us explain this embedding on a small illustrative example where a part $m$ of the message has to be embedded into this x LSB of pixels, which are respectively a 3 bits column vector and a 7 bits column vector. Let then $H$ be the binary Hamming matrix

\[ H = \left( 
\begin{array}{lllllll}
	0 & 0 & 0 & 1 & 1 & 1 & 1 \\ 
	0 & 1 & 1 & 0 & 0 & 1 & 1 \\ 
	1 & 0 & 1 & 0 & 1 & 0 & 1
\end{array}\right)
\]

The objective is to modify $x$ to get $y$ s.t. $m = H y$. In this algebra, the sum and the product respectively correspond to the exclusive \emph{or} and to the \emph{and} Boolean operators. If $H x$ is already equal to $m$, nothing has to be changed and $x$ can be sent. Otherwise we consider the difference  $\delta= d(m,H x)$ which is expressed as a vector:

$$\delta = \left( 
\begin{array}{c}
    \delta_1 \\ 
	\delta_2 \\ 
	\delta_3
\end{array}\right) ,$$
where $\delta_i$ is 0 if  $m_i = H_{x_i}$ and 1 otherwise.

Let us thus consider the $j$-th column of $H$, which is equal to $\delta$. We denote by $x_j$ the vector we obtain by switching the $j$-th component of $x$, that is,  $\overline{x^j}=(x_1; \ldots; \overline{{x}_j}; \ldots ; x_n)$. It is not hard to see that if $y$ is  equal to $\overline{x^j}$ , then $m = H_y$. It is then possible to embed 3 bits in only 7 LSB of pixels by modifying on average $1-2^3$ changes. More generally, the F5 embedding efficiency should theoretically be $\frac{p}{(1-2^p)}$.

However, the event when the coefficient resulting from this LSB switch becomes zero (usually referred to as shrinkage) may occur. In that case, the recipient cannot determine whether the coefficient was -1, +1 and has changed to 0 due to the algorithm or was initially 0. The F5 scheme solves this problem first by defining a LSB with the following (not even) function:

\[ LSB(x) = \left\{ 
\begin{array}{ll}
	1-x & \textnormal{ mod 2 if $x<0$}, \\
	x & \textnormal{ mod 2 otherwise}
\end{array} 
\right. \]
Next, if the coefficient has to be changed to 0, the same bit message is re-embedded in the next group of $x$ coefficient LSB.

The scheme nsF5 focuses on steps of Hamming coding and ad hoc shrinkage removing. It replaces them with a wet paper code approach that is based on a random binary matrix. More precisely, let $D$ be a random binary matrix of size $x × n$ without replicates nor null columns: consider for instance a subset of $\{1,2^x\}$ of cardinality n and write them as binary numbers. The subset is generated thanks to a PRNG seeded with a shared key. In this block of size $x$, one choose to embed only $k$ elements of the message $m$. By abuse, the restriction of the message is again called $m$. It thus remains $x- k$ (wet) indexes/places where the information should not be stored. Such indexes are generated too with the keyed PRNG. Let $v$ be defined by the following equation: $$D_v=\delta(m,D_x).$$

\begin{figure}[ht]
\begin{center}
  \subfigure[t=50]{\includegraphics[scale=0.45]{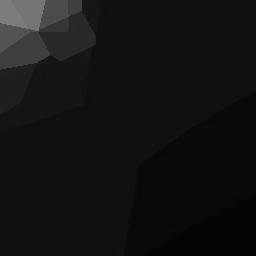}}\quad
  \subfigure[t=75]{\includegraphics[scale=0.45]{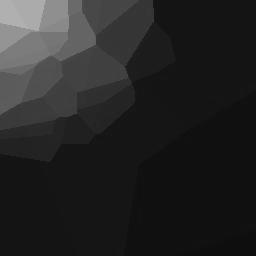}}\\
    \subfigure[t=90]{\includegraphics[scale=0.45]{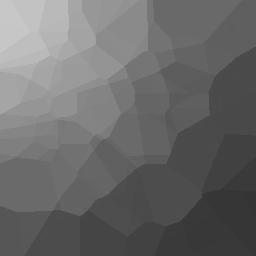}}\quad
  \subfigure[t=100]{\includegraphics[scale=0.45]{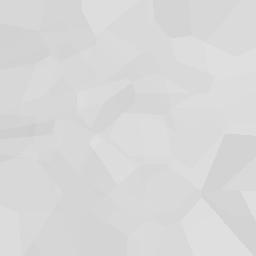}}
\end{center}
\caption{The wireless sensor network at various dates}
\label{fig:WSN}
\end{figure}

This equation may be solved by Gaussian reduction or other more efficient algorithms. If there is a solution, one have the list of indexes to modify into the cover. The nsF5 scheme implements such an optimized algorithm, that is to say, the LT codes.
 
Let us now apply this algorithm in order to modify locally the sensed 
data without being detected.
 
\subsubsection{Simulation protocol and results}

In this set of experiments, the sensors network computed using
Python language is
constituted by $256^2$ individuals, sensing respectively
the temperature (50 \% of the sensors), 
pressure (40 \%), and humidity (10 \%) levels on the 
area under consideration. 
Moreover, the physical measure evolution is defined as follows.
100 particular homogeneous areas have been defined on the monitored place,
on which temperature (50 remarkable locations), pressure (40), or humidity (10 locations)
are constant. 
We then have supposed that at time $t$ and location $(x,y)$, and 
after normalization:
\begin{itemize}
\item temperature follows a Gaussian law of
parameter $(40 (1+0.005 t/4 \sqrt{x^2+y^2}), 5)$;
\item the Gaussian parameters are 
$(40 (1+0.01 t/4 \sqrt{x^2+y^2}), 5)$ 
for pressure;
\item finally, the 10 humidity remarkable locations produce 
data following a Gaussian law of parameter 
$(40 (1+0.001 t/4 \sqrt{x^2+y^2})$.
\end{itemize}
At each location $(x,y)$, a color pixel is associated
to the sensed value, temperature being its red component while
pressure and humidity are respectively associated to green and blue.
Examples of image-represented data produced by the wireless
sensor network are represented in Figure~\ref{fig:WSN} after gray scale 
conversion.

\begin{figure}[ht]
\begin{center}
  \subfigure[t=50]{\includegraphics[scale=0.45]{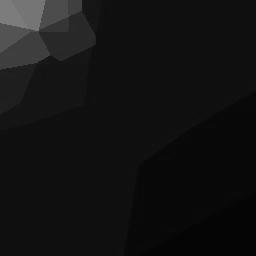}}\quad
  \subfigure[t=75]{\includegraphics[scale=0.45]{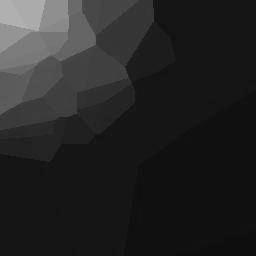}}\\
    \subfigure[t=90]{\includegraphics[scale=0.45]{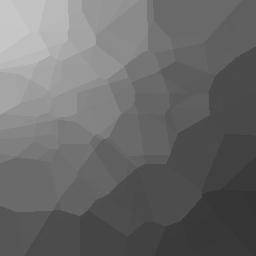}}\quad
  \subfigure[t=100]{\includegraphics[scale=0.45]{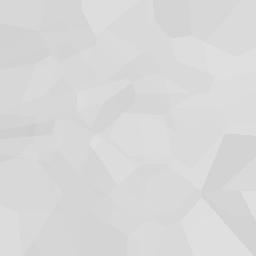}}
\end{center}
\caption{The attacked wireless sensor network at various dates}
\label{fig:WSN2}
\end{figure}

\begin{figure}
\begin{center}
  \subfigure[Out-of-bag error]{\label{roc1}\includegraphics[scale=0.38]{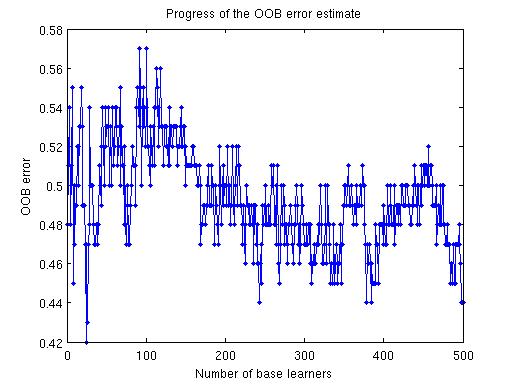}}~~
  \subfigure[Optimal subspace functionality]{\label{roc2}\includegraphics[scale=0.38]{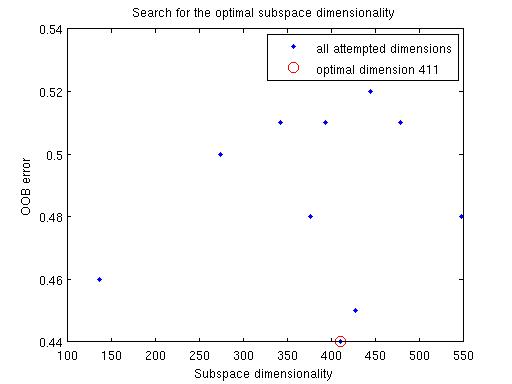}}\\
    \subfigure[ROC curve]{\label{roc3}\includegraphics[scale=0.5]{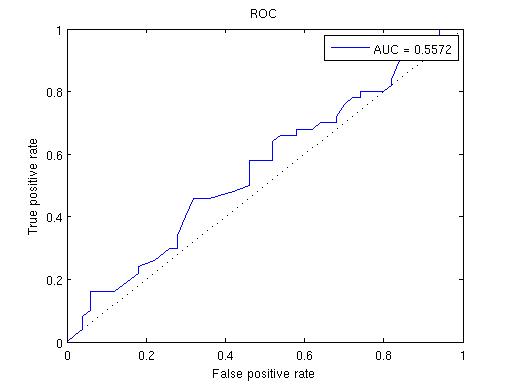}}
\end{center}
\caption{ROC curve and OOB error at sink level}
\label{fig:roc}
\end{figure}

An attack has then been realized on the area under surveillance
on locations (physical attack of sensors) provided by the nsF5~\cite{nsF5} algorithm 
with a payload equal to 0.1 bits per non-zero AC DCT coefficient.
Figure~\ref{fig:roc} contains the receiver operating characteristics
ROC curve obtained at the sink level, using steganalyzer published in~\cite{6081929}. 
This classifier uses the features extracted by the method called 
CC-PEV~\cite{Kodovsky}, and it divides the collected features into two sets (training set and testing set). This ensemble classifier is indeed composed of many base learners, each base learner is trained on sets
 $\{{X_m}$, ${{\overline{X}}_{m}}\}$
for cover and stego features. Each base learner is implemented as the Fisher Linear Discriminant. It is independently trained on the data composed by
cover and stego features. The final decision is obtained by assembling
each result produced by these base learners. As it
can be observed, the obtained ROC curve is close to the first diagonal, leading
to the conclusion that the area observed by both natural and faked sensors
would probably appears as natural for the sink.
The testing error called \emph{out-of-bag}  (OOB)  it  estimated 
using equation~\ref{eqref1}. 
\begin{equation}
\label{eqref1}
E^{(n)}_{OOB}=\frac{1}{{2N}^{trn}}\sum_{m=1}^{{N}^{trn}}({B}^{(n)}({X}_{m})+1-{B}^{(n)}(\overline{X}_{m}))
\end{equation}
in which $X_m$ and $\overline{X}_m$  are respectively features of original and
associated stego features, $m$ $=1,...,N^{trn}$, for the cover and stego feature vectors, $n$ is the number of trained base learners, $trn$
is the size of the training set.
Figures~\ref{roc1}, \ref{roc2} show the progression
of the error with the number of base learners. 
As it can be seen, this error does not significantly decrease when the number
of base learners increases.
It can be deduced from these results that the adversary has
achieved to slightly modify some sensors without being detected
at sink level.

\section{Conclusion}
\label{sec:conc}


In this article, the usefulness of information hiding technologies for various WSN
security concerns has been evoked. It has been described in which contexts steganographiers,
steganalyzers, fragile or robust watermarking schemes can be used, and the associated 
security concern has been detailed. Furthermore, an original application of a well-known
steganographier, namely the nsF5 algorithm, has been experimented, while the interest of
steganalyzers to detect malfunctioning observed devices has been signaled.

In future work, the authors intention is to investigate more deeply the links evoked between
WSN and IH techniques. Cryptographic definitions of security in steganalysis domain will be
reformulated in terms of WSN security. A WSN specific malfunctioning detector based on
steganalysis literature will be designed and tested, while an attack on a real network 
already deployed will be executed using the nsF5, to prove the effectiveness of the approach.

\bibliographystyle{plain}
\bibliography{biblio}
\end{document}